# Antecedents for Successful Collaboration in Requirements Engineering

**Risto Paavola**
Aalto University School of Economics
Helsinki, Finland
Email: risto.paavola@gmail.com

**Petri Hallikainen**
The University of Sydney Business School
Sydney, Australia
Email: petri.hallikainen@sydney.edu.au

## Abstract

The main focus of the requirements engineering (RE) literature has been on the technical aspects related to the RE projects. Furthermore, research has mainly focused on the specific methods for collecting the requirements for an information system. To fill this gap, this paper studies the contribution of social factors, such as social ties, knowledge sharing and flexibility, to successful collaboration in RE teams. Data were collected from a successful RE and development project in a public sector company in Finland. The results suggest that human-related issues, such as flexibility and transactive memory, were important for collaborative work in the RE team studied. The paper concludes by discussing the implications for theory and suggesting practical guidelines to enhance collaborative work in RE teams.

## 1   Introduction

The most critical part of information systems development (ISD) process is requirements engineering (RE) (e.g. Schenk et al. 1998). A significant portion of ISD failures are attributed to incomplete and inaccurate information requirements (Byrd et al. 1992, Vessey and Conger 1993, Watson and Frolick 1993, Wetherbe 1991). Incorrect requirements cause cost and schedule overruns in ISD projects (Vessey and Conger 1994). It is crucial to conduct as complete as possible requirements for an information system since the cost of modifying the software after implementation or at a later stage in the development process is considerably higher than that of making changes earlier in the development process (Shemer 1987).

RE is not, however, an easy task. It has been acknowledged in the literature that RE is an iterative learning process (e.g. Majchrzak et al 2005, and Xu and Balasubramaniam 2007). Starting with high level requirements the RE process proceeds to more and more detailed requirements. In this process unexpected problems, such as issues related to technology or business processes, are likely to occur. Because RE is essentially iterative and a learning process, it is not possible to foresee all the potential issues, nor is it possible to determine early in the process the individuals needed to solve those issues. Against this backdrop, we propose that the interaction between people who influence and are influenced by the RE process is crucial to achieve success.

The existing literature is rather silent about the social interaction related to the RE processes and the role of the social network in the RE process. Earlier research has acknowledged that the knowledge and experience of the team members is important for RE (Liou and Chen 1993). Aspects, such as team creativity, communication, motivation and individual skills of the team members have also been discussed in the existing literature (Tiwana and McLean 2005). However, in the present research we submit that we should focus the attention beyond the RE team and to the wider social network.

There is a rather large body of literature on the formal techniques and methods for RE. Some of the techniques developed include: asking and deriving from existing information system, strategy set transformation, decision analysis, socio-technical analysis, interrogatories technique, and a semantic questioning scheme (Ackoff 1967, Appan and Browne 2012, Bostrom and Heinen 1977/1, Bostrom and Heinen 1977/2, and King 1978). Formal methods presented in the literature included, for example, analyst driven requirement engineering, and collaborative requirement engineering (e.g. JAD).

Although it is acknowledged that stakeholder collaboration is essential for RE, we have not found literature focusing on the informal social interaction within the RE process.

The objective of the present research is to investigate the social ties and knowledge sharing in the social network within an RE process and how they affect successful collaboration in the RE process. We draw from the model for successful collaboration in virtual teams by Kotlarsky and Oshri (2005) to form the theoretical foundation for our study. Although their model was tested in a slightly different context, we propose that knowledge sharing and social ties are relevant factors also in the context of RE. A case study where all the relevant participants in the RE process were interviewed is presented in this paper.

The results show that particularly transactive memory was considered important for successful collaboration in the RE process. Additionally, we discovered that "flexibility" had great importance to a successful RE process. It was evident that flexibility in terms of making compromises to reach an acceptable solution for all parties was very important for reaching a successful outcome.

The article is organized as follows. The next section reviews the existing literature on RE and section three briefly describes the theoretical foundations for analyzing social ties and knowledge sharing in the RE process. Section four explains the empirical research methodology and the findings are presented in section five. Finally, the results are discussed and the paper is concluded in section six.

## 2 REQUIREMENTS ENGINEERING

Information Systems Development (ISD) is inherently a creative process because it involves generation and evaluation of new ideas, designs, solutions, and artifacts (Tiwana and McLean 2005). The tasks in the ISD process are knowledge intensive and involve multiple stakeholders (Krauft and Streeter 1995), which can result in several issues in the process. When requirements are elicited from a heterogeneous group, communication breakdowns are likely to occur (Liou and Chen 1993). Cognitive and communication issues generally make the RE process challenging (Browne and Rogich 2001). Requirements are hard to define because humans are characterized by limited memory, bias in selection and use of data, and bounded rational problem-solving behavior.

The RE phase of the ISD process answers to three major questions: why (the problems and the needs being addressed), what (requirements), and how (conceptual design) (Shemer 1987). The notion of requirements uncertainty has been the focus of research for decades (Lyytinen et al. 1998). Project risk researchers have identified a number of project risk drivers that are related to requirements (e.g. Barki et al. 1993; Schmidt et al. 2001). RE risks are mostly associated with organizational issues, individual competence and team dynamics (Li et al. 2011, Wallace et al. 2003, and Xu and Balasubramaniam 2007).

According to Byrd et al. (1992) and later Browne and Rogich (2001) requirements are organized in four different levels: goal, process, task and information. Goal level requirements set the organizational goals for the system under development. Process level requirements describe business activities and tools, which are derived from the goal level requirements. Task level requirements describe specific steps to fulfill the activities described in the process level requirements. Information level requirements describe the data needs and relationships and they are derived from the previous levels. There are several techniques to gather requirements. The simplest forms are asking and deriving from existing information system. Advanced strategies are for example: strategy set transformation (King 1978), decision analysis (Ackoff, 1967), and socio-technical analysis (Bostrom and Heinen 1977/1 and Bostrom and Heinen 1977/2). Information system users' requirements can be guessed when information is pushed to the users or requirements can be elicited from users (Sun 2003). End users have the knowledge of the current company processes (Muller et al. 1993). Moreover, they can describe the culture of the company and are able to see both the flaws and the best practices in the current processes. The change of the business processes must be "designed" in the RE phase (Schmidt et al. 2001) to identify the required business changes and to get the users on board. If there is no or little application domain knowledge, most of the time needs to be spent on learning in the RE phase (Walz et al. 1993).

RE is often characterized as an ongoing sense-making process among stakeholders that is chaotic, non-linear and continuous (Walz et al. 1993). To overcome this, several software process standards or reference models defining software development process stages and roles have been developed, such as ISO/IEC 12207, IEEE/EIA 12207, and Rational Unified Process® (RUP®). The single solution for all thinking embedded in these models has been criticized by researchers (Jiang et al. 2009). Slaughter et al. (2006) suggest that software processes need to be customized for different types of products and business strategies. Thanasankit (2002) emphasized that aspects of national culture, such as the perception of the concept of power can change the way RE processes are implemented in organizations. A lot of research has been conducted on how to choose the right RE technique, such as Hickey and Davis' unified model of requirements engineering (2004), but much less attention has been given to the role of the requirements engineering team.

The RE process is a collaborative effort among managers, users, and system analysts (Liou and Chen 1993). Wetherbe (1991) suggests that requirements engineering is a cross-functional, joint application design that involves input from all key decision makers involved in a business process. The RE team naturally plays a central role in the process. Team creativity is inherently a social process that builds on and incorporates individual creative processes at the project level (Tiwana and McLean 2005) and results from individual knowledge, cognitive factors, and task motivation as well as from group level tasks, norms, diversity, and problem solving (Cooper 2000). Individual talent and experience reduces task or technology related risks (Henderson and Lee 1992). Knowledge is the raw material and problem solving is the core of RE. Knowledge from multiple technical and functional domains is a necessity for RE (Curtis et al. 1988).

RE is not an easy task since there are several social factors such as cognitive and communication issues (Browne and Rogich 2001). Curtis et al. (1988) classified requirement risks to the thin spread of application domain knowledge, fluctuating and conflicting requirements, and communication and coordination breakdowns. Unwanted conditions arise from unclear goals and objectives, which cause difficulty for accurate project scope determination (Hickey and Davis 2004). Furthermore, individual team members often have their own partial mental models about the design problem and possible solutions, which are biased by their prior experience in other ISD projects (Robillard 1999).

Commitment of different stakeholders is crucial to the success of the ISD project (Newman and Sabherwal 1996). Newman and Sabherwal (1996) identify four types of determinants that affect on commitment: project, psychological, social, and structural. The difference in perceptions among the people involved in an ISD project is called equivocality. It limits the mutual understanding needed to accomplish the goals of the project (Daft and Lengel 1986). In the present paper we propose that knowledge sharing and social ties between the stakeholders are very important for reaching the mutual understanding on the requirements. The next section describes the theoretical foundations to study the role of knowledge sharing and social ties in the RE process.

## 3 SOCIAL COLLABORATION ASPECTS IN REQUIREMENTS ENGINEERING

Kotlarsky and Oshri (2005) studied the collaboration in virtual teams using a research model describing knowledge sharing and social ties as factors affecting successful collaboration in virtual teams. We draw from their model to establish the theoretical foundation for our study. Although the context of virtual team work is different from the RE process, we believe that the model is applicable for our research because of the reasons outlined below.

Knowledge in organizations is created and maintained within organizational units, groups or communities of practice. Because of the specific nature of knowledge it is difficult to move knowledge across these communities (Brown and Duguid 1998). As the RE process requires knowledge from different stakeholders located in different organizational units with specific knowledge the problem of knowledge sharing is likely to exist in the RE processes as well. It is thus reasonable to assume that the ability to share knowledge would favorably affect the successful collaboration in the RE process. We follow Kotlarsky and Oshri (2005) and divide knowledge sharing to the two categories of transactive memory and collective knowledge. Transactive memory means that the individuals should rather know who has the expertise than try to gain all the expertise to themselves (Wegner 1987). Collective knowledge is defined as "tacit and collective; a knowledge of the unspoken, of the invisible structure of a situation, a certain wisdom that is acquired through social practice" (Baumard 1999, p. 66).

As the participating stakeholders in the RE process most likely have differing objectives and different agendas, successful collaboration requires trust between the participants (Child 2001). Moreover, social and cultural aspects affect the RE process (Hanisch et al. 2001). It has been shown that informal personal relationships are needed for trust to develop (Arino et al. 2001). We follow Kotlarsky and Oshri (2005) and propose that the social ties consisting of trust and rapport are important factors affecting the successful collaboration in the RE process. Trust is defined as "the willingness of one person or group to relate to another in the belief that the other's action will be beneficial rather than detrimental, even though this cannot be guaranteed" (Child 2001, p. 275). Rapport is defined as "the quality of the relation or connection between interactants, marked by harmony, conformity, accord, and affinity' (Bernieri et al. 1994, p. 113).

## 4    RESEARCH METHOD AND APPROACH

An in-depth case study of an information system requirements engineering project is presented in this paper. The case study method is suitable for researching complex phenomena in their real life contexts (Yin 1994, p. 13). As recommended by Yin (1994), multiple data sources were used in this study: interviews, emails, and project documents (Eisenhardt 1989; Yin 1994). As the phenomenon of interest for the research was defined in advance and we applied an existing conceptual model and initial categories and codes in our analysis, the research approach can be classified as an instrumental case study (Stake, 1995).

One of the researchers was the Chief Information Officer (CIO) in the case company and acted in this role in the case project. The purpose was, however, not to conduct action research, but rather to observe and understand how the requirements engineering team performed in the case project. The study thus falls between the soft and hard case study approach according to Braa and Vidgen's (1999) case research framework. We see the participation of the researcher in the case as a strength because it enabled richer access to data. The researcher participated in the project in his professional role and only recorded his observations for research purposes.

Davidson (2002) proposed that requirements development should be recognized as a socio-cognitive process. The purpose of this study is to deepen the understanding of the requirements engineering phase of information systems development as a complex socio-political process. This research studies the ways in which these processes are played out in actual organizations and over time. Thus, this research has an ensemble view of technology (Orlikowski and Iacono 2001) and it studies technology as a development project (Sidorova et al. 2008). We adopt the socio-technical perspective to research the building of an IT artefact by a project group. The units of analysis in this research are the project team and the members of the team so this research uses a mixed-level unit of analysis. A mixed-level strategy preserves the macro-level concepts and grounds these concepts on individual purposes and behaviour (Markus and Robey 1988). Theory is used as a lens for characterizing and analysing the ICT project at hand.

### 4.1    Data Collection

The snowballing method (Goodman 1961) was used to identify the members of the social network in RE process. All the interviewees were asked who they were working with and how in the RE process and strong and weak links between the participants were identified from this data. All the RE participants who had strong links to each other were interviewed for this research and are presented in Figure 1.

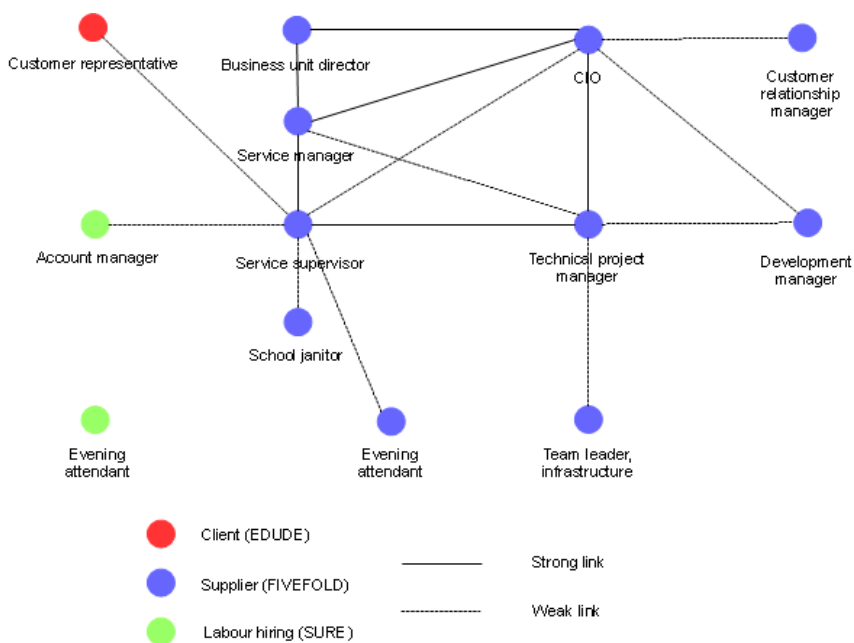

*Figure 1. Social Network of the RE Team in the Case Project*

The research interviews were conducted after the production usage was started. In the interviews, focused questions were asked from the following areas: 1) the participants and their responsibilities of the RE process, 2) the process of RE, 3) the achieved change, and 4) the results of the ISD project. The questions were open ended and additional information was asked if the respondents gave new or interesting information, or if they gave incomplete information. The interviews were recorded and notes were made during the interviews by the researcher. After transcribing the interviews it was noticed that there were some unanswered questions and contradicting answers. Answers to these questions were directly acquired via email from the corresponding individual. All the relevant RE team members were interviewed twice: director of business unit (business dept.), service manager (business dept.), service supervisor (business dept.), and technical specialist (ICT dept.). The CIO (ICT dept.) could not be interviewed since he was one of the researchers.

To triangulate the findings, direct observations from the process (written down as memos) and informal interviews were made by the researcher. Also meeting notes, emails and project documentation were used as data sources. The data finally consisted of about 7.5 hours of interview recordings, 100 pages (A4) of interview transcriptions, 39 pages (A4) of interview memos, 65 pages of emails, 14 pages of meeting memos, 8 pages of technical documentation, and 9 pages of researcher's notes and all of it was in Finnish.

We used the model of Kotlarsky and Oshri (2005) to establish a set of initial categories to be used in the analysis (Miles and Huberman, 1994, p. 61). We were open for additional categories to emerge throughout the coding process and the codes were revised as the analysis proceeded. The process of the coding is presented in Figure 2. The coding was performed using Google Drive's Google Sheets.

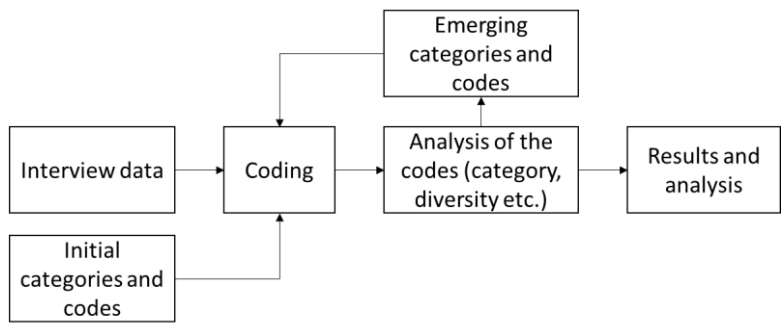

*Figure 2. The Coding Process for the Research*

The codes were identified from chunks of text that are partial or complete sentences or expressions describing specific activities (Strauss and Corbin 1998). Codes were associated with categories. A bottom-up interpretive approach was used to associate new codes with particular categories and concepts. In the coding process the researchers found new codes which didn't quite fit to the initial categories. They were coded, collected, analysed, and categorized. Thus, a new concept emerged: "flexibility", which consists of codes such as compromising, complying, and having a sense of direction when needed. In the statements analysed, codes were identified and grouped, and their association with categories as well as their corresponding concepts were established. For example, in a sentence "I think that at the end the security policies were working, because I didn't get exactly what I wanted first, because I didn't understand all the security aspects… The project was a great success" the chunk "I didn't get exactly what I wanted first" describes code "compromising" (in category flexibility) and "The project was a great success" describes code "successful product" (in category successful collaboration). Causalities between the concepts were categorized as strong or weak. A strong link was identified if the interviewee directly told that a certain identified phenomenon (code) caused another. A weak link was identified if the interviewee told, for example, at the beginning of the interview that the project was a success and then later on described how social trust was achieved between the team members (in this example trust causes product success, but the causal relationship is considered as weak).

## 4.2 Case Project

FIVEFOLD (a pseudonym) is a large Finnish public sector commercial enterprise. It provides catering, property maintenance, cleaning, security, and telephone and wellness services for a large city. FIVEFOLD employs approximately 3000 workers.

The case under study was an information system development project of the property maintenance department's public owned schools' janitoring process. There were approximately a hundred schools and their premises are owned by a bureau of the city called EDUDE (a pseudonym). The school premises are offered for renting to sports clubs and private citizens, when the premises aren't used for teaching purposes. FIVEFOLD's responsibility was to provide employees to the school premises for surveying the premises. There were only a couple of FIVEFOLD's own employees assigned since most of the employees were rented from a labour hire company called SURE (a pseudonym). In terms of information management, the most important data for all the companies were the ordered rental hours by the renters using the premises and the tracking of the hours worked by the employees. The end users (renters) booked the school premises beforehand and were invoiced by EDUDE for the hours they were using the school premises. EDUDE offered the end users (the renters) an ICT system for school premises booking. SURE employers utilized SURE's system for keeping track of their working hours. SURE invoiced FIVEFOLD for the hours and FIVEFOLD invoiced EDUDE. There was some uncertainty in terms of the usage hours in situations when renters didn't arrive or overused their rental time. Because of the complicated service chain, dispersed non-integrated ICT systems and undecided master data management, the companies involved in school premise renting suffered from low quality and dispersed data. The poor quality of data caused invoicing problems and assumedly misuse of the system for tracking working hours. These issues were the background for starting the project described below.

The very first initiative to the new ICT system development project was made by FIVEFOLD's service manager in summer 2013, who discussed with FIVEFOLD's CIO about the problems and the need for a new information system. The schedule was rather tight since the business would start along with the school semester so there was about three months available for development and at the same time the summer vacations were still on. The requirement engineering had actually been started beforehand, when the service supervisor contacted the CIO unofficially in spring 2013. In the beginning of June the service supervisor and the CIO of FIVEFOLD continued to narrow down the problems in the current business process and at the same time gathering the requirements for the new system. The existing invoicing system was based on Excel. Invoicing in FIVEFOLD was conducted by using 144 Excel sheets located in a shared folder on a network drive. The service supervisor collected the working hours of employees from SURE's time tracking system and input them on an Excel sheet and based on its calculations wrote an invoice to EDUDE. There were complicated calculation formulas in the Excel sheets, which were prone to errors. One copy-paste error could cause a several tens of thousands mistake in an invoice.

The service supervisor and the CIO discussed about the master data management. Based on previous discussions between the service supervisor and the client representatives from EDUDE, integrating the planned hours from EDUDE's information system was impossible. The service supervisor had also been discussing with SURE's representatives and integrating the working hours from SURE's time tracking system was also considered impossible. It was a difficult situation for FIVEFOLD to be responsible for a human produced business when the planned hours were input into the customer's ICT system and the realized working hours were input into the outsourcing partner's ICT system. The idea was that there should be a new information system, which collected the working hours from the SURE employees and compared them to the booked hours. The system should report the hours that were done as planned and pinpoint those which weren't done according to the plans. The service supervisor proposed that SURE's employees would input the hours, FIVEFOLD's janitors (in school premises) would accept the hours, and the service supervisor would invoice the hours. This was realized to be a problem because workstations in schools were meant only for the employees of FIVEFOLD and they were connected to a private network and domain. FIVEFOLD had a mobile ERP system which was considered but was turned down by the service supervisor because of the license costs and the business unit's bad experiences from the past. FIVEFOLD had also a basic LAMP (Linux, Apache, MySQL and PHP) codebase of the time tracking system, which could be modified to the needs of the business. However, the service supervisor suggested an Excel based system.

Shortly after meeting the service supervisor, the CIO met with the FIVEFOLD director of the business unit. In the meeting, the director of the business unit presented the contracts and the invoicing process. The CIO presented his ideas and discussed the risks of the solutions. The director explained the issues related to the service contracts between the parties. The differences between arrangements led to a situation that some hours were profitable for FIVEFOLD, but some were not. The service was barely profitable and thus the costs of the ICT tool couldn't become very high. The director of the business unit described that the process was almost completely manual.

In mid-July the CIO of FIVEFOLD assigned a technical specialist from the company's ICT department to work in the project. The technical specialist's job was to continue eliciting requirements and developing the tool in collaboration with the business unit. It was agreed that it was a business decision whether to use sheet based systems but the risks of it should be clearly communicated to the business unit from the ICT department. It was decided to conduct a market survey of commercial time tracking tools. The technical specialist used mainly interviews to gather the requirements from the different parties. She wrote the first version of the requirements specification document in mid-July and presented it to the CIO. The CIO added the technical requirements so that the system could fit to the corporate ICT architecture.

The CIO and the technical specialist agreed on developing the tools to the demo phase and presenting all the alternatives to the business unit, which could then make the business decision. In late July the technical specialist presented the first version of the sheet based information system to the CIO. The CIO made some improvement notions. The CIO presented the current progress of the LAMP tool. The technical specialist tested the tool and found some errors, which were corrected "on the fly". The CIO initially wanted the tool to be integrated to the corporate invoicing system, but eventually they both agreed that this integration was costly and would require 3rd party developers. It was decided that the feature should not be developed in the demo phase.

The technical specialist met the service supervisor in the end of July. It was decided that the technical specialist continued gathering the requirements in collaboration with the service supervisor. New requirements came up and the new business process was documented in more detail. The technical specialist presented the demos to the service supervisor, who adapted the requirements documentations based on them. In the beginning of August the CIO decided to stop the development of the LAMP based system since there were already enough features for the demo purposes and the decision had not been made yet so it was not clever to sacrifice significant amount of additional resources to that. At the end of July the technical specialist met with the service manager and she presented him the alternatives and the gathered requirements. The service manager made a decision to go with the sheet based information system. The most important decision criterion was the tight schedule; the information system needed to be up & running at the beginning of September. The technical specialist and the CIO met and documented the risks that the decision included.

A meeting was organized between the service manager and the CIO. According to the CIO the company should acquire 3rd party software and its support services and tailor the software to fit the business unit's needs. The service manager noted that the business has so low profitability that a lot of money cannot be spent. The CIO presented the service manager the demo of the LAMP based system. The service manager decided that due to the tight schedule, the business unit would use the sheet based system and the development of the LAMP based system should continue and it could be taken into use at the end of the year during the holidays.

The sheet based tool was developed in two week cycles. The technical specialist developed the system by herself. There was no need to write a single line of code, just to use the existing systems in different ways. At the end of each cycle, the technical specialist presented the tool to the service supervisor, who gave comments and new requirements if necessary. Testing and documentation were done simultaneously. Eight of FIVEFOLD's own evening attendants were the pilot group and were trained to use the system.

In the end of August, the project faced problems. It was only a couple of days until the launch when it was noticed that SURE employees could not access the developed sheet based system with their shared user accounts. There was no time left to create personal external user accounts for all SURE employees. The business unit made an emergency decision that SURE employees wouldn't input the hours to the system but the FIVEFOLD school janitors would do it. It was mid-September at the time and the production use had already started in the beginning of September.

The system was presented to the customer (EDUDE), which immediately presented new requirements. These functionalities were developed shortly after that. EDUDE started using the system according to their needs. New requirements started coming also from the end users. The system was "trimmed" according to the comments of FIVEFOLD school janitors using the system on the daily basis.

Overall the project was successful keeping in mind that the user base was large and the schedule was tight. The service supervisor felt that the developed system helped and hastened the work and the data collected was mostly accurate and correct. After the project the business process became more straightforward and there were less error prone tasks. The business unit was very satisfied with the continuous support from the ICT department. The system even brought some unexpected benefits: some school janitors use the system to collect data on the work introduction of evening attendants.

## 5   RESULTS

Based on the empirical evidence presented below, we argue that social ties and knowledge sharing contributed to successful collaboration in the company and the case studied.

In order to support the above claim we followed the methodology from Kotlarsky's and Oshri's study (2005) and had three layers of analysis. Three levels of evidence will be outlined in the following section. The first level is an outline of statements made by interviewees associated with the concepts under investigation. The second level is the frequency of these statements. The third level will present the number of instances in which social ties, knowledge sharing, and flexibility (an additional category, which emerged during the coding process) were linked to successful collaboration.

**Social ties in the requirements engineering team: evidence**

Statements made by interviewees about rapport and trust are presented below. These statements were analyzed and associated with rapport and trust based on the definitions provided in Kotlarsky's and Oshri's study (2005).

**Rapport**: *"Tuija [technical specialist] also thinks about different solutions and choices and is generally open towards me, tells me the choices and gives her opinion, which should be chosen."* (Satu)

**Trust**: *"[Johannes'] knowledge was good enough for his role. He trusted me and my knowledge."* (Satu)

**Knowledge sharing in the requirements engineering team: evidence**

Statements made by interviewees about transactive memory and collective knowledge are presented below. These statements were analyzed and associated with transactive memory and collective knowledge based on the definitions provided in Kotlarsky's and Oshri's study (2005).

**Transactive memory**: *"Then we [business unit] wanted [technical] specialist's opinion, what the solution would be."* (Satu)

**Collective knowledge**: *"I [Satu] and Tuija never talked about roles [in the RE project] it [work and responsibility sharing] just happened."* (Satu)

**Flexibility in the requirements engineering team: evidence**

A statement made by an interviewee about flexibility is presented below. These statements were analyzed and associated with flexibility based on the new concept, which emerged from the data.

**Flexibility**: *"First you [IT department] came up with this high end solution, which was overkill for this problem. But then it didn't work out and we did [together] this good enough solution."* (Juha)

**Successful collaboration in the requirements engineering team: evidence**

Successful collaboration was defined by the perception of interviewees that a project team was collaborative, the RE process succeeded, or the product produced was a successful one. Successful product indicators can also be objective, such as lower workload, more straightforward business process, or less errors in the process. The perceptions of the interviewees with regard to product success and personal satisfaction, representing successful collaboration are presented below. These statements were analyzed and associated with product success and personal satisfaction based on the definitions provided in Kotlarsky's and Oshri's study (2005).

**Product success**: *"The end result [of the project] was wonderful and unbelievable."* (Satu)

**Personal satisfaction**: *"Now I really get good support [from IT department] and collaboration is superb. Everything works for me."* (Johannes)

**Product and project success (objective evidence)**

After the project the business process was more straightforward and there were less error prone tasks. For example, before the project school janitors got the time usage exceptions as paper notes, which they copied and mailed (physically) to both EDUDE and FIVEFOLD invoicing. After the project, the janitors just input the hours and the exceptions to the system and EDUDE and FIVEFOLD representatives could go there and check the markings. This led to the situation that the workload of the service supervisors dropped to about 10 % from what it used to be. The invoicing process was also improved. The invoices from the beginning of January were created in time, were accurate and the customer (EDUDE) didn't complain about them. The system also brought some unexpected benefits. It was noted that some school janitors used the system to collect data from the work introduction of evening attendants, which helped the HR department of the company.

**Concept frequencies for social ties, knowledge sharing, flexibility and successful collaboration**

In all, 127 statements were made by interviewees that were associated with codes related to these concepts. Moreover, 'Diversity in codes' was calculated, which represents the number of different codes grouped within one category. For example, under the category 'trust' four different codes were identified. Our calculations in Table 1 show that 23 statements were made with regard to social ties, 47 statements concerning knowledge sharing, and 18 statements with regard to flexibility. Within the concepts, a large number of statements were associated with transactive memory (31). These findings may suggest that interviewees have considered social network and its knowledge to be an important element in collaborative RE work. The importance of social ties and knowledge sharing in successful collaboration will be further discussed in the following section.

*Table 1. Concept Frequencies Based on the Number of Statements*

| Concept | Categories in concept | Diversity in codes | Concept frequencies |
| --- | --- | --- | --- |
| Social ties | Rapport | 10 | 15 |
| | Trust | 4 | 8 |
| Knowledge sharing | Transactive memory | 10 | 31 |
| | Collective knowledge | 13 | 16 |
| Flexibility | Compromising | 8 | 18 |
| Successful collaboration | Product success | 17 | 28 |
| | Personal satisfaction | 5 | 10 |

**The relationships between social ties, knowledge sharing, flexibility and successful collaboration**

To assess the importance of social ties and knowledge sharing for successful collaboration, a calculation was made in a similar fashion as Kotlarsky's and Oshri's study (2005), with the exception that strong and weak causalities were identified and calculated separately. The number of statements that represented explicit relationships (strong causalities) between social ties, knowledge sharing, flexibility, and successful collaboration were calculated. These calculations are presented in Table 2 under the columns '[Strong/Weak] relationships with successful collaboration'. Two conclusions can be drawn from the calculations presented in Table 2. Firstly, Table 2 suggests that social ties, knowledge sharing, and flexibility were positively associated with successful collaboration in 78%, 79%, and 94% (sums of strong and weak causalities) of the statements made, respectively. The causality found is described more closely in Figure 3. Secondly, flexibility had the strongest relationship with successful collaboration both in terms of strong and weak causality. Based on the evidence above, we argue that our findings suggest that human-related issues in the form of social ties, knowledge sharing, and especially flexibility were considered as the key to successful collaboration in RE.

*Table 2. Calculated Values of Relationships Between Concepts Based on Number of Associated Codes*

| Concept | Concept frequencies (count from Table 1) | Strong relationships with successful collaboration (statements/per cent) | Weak relationships with successful collaboration (statements/per cent) |
| --- | --- | --- | --- |
| Social ties | 23 | 9 (39%) | 9 (39%) |
| Knowledge sharing | 47 | 16 (34%) | 21 (45%) |
| Flexibility | 18 | 9 (50%) | 8 (44%) |

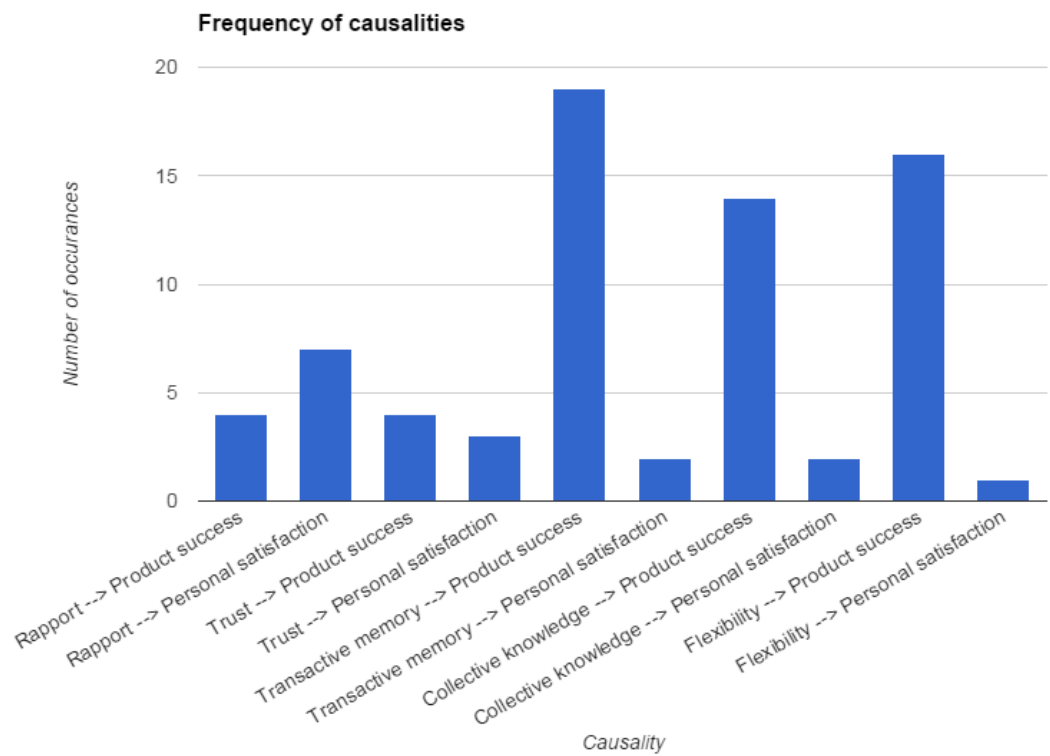

*Figure 3. The Frequency of the Category Causalities*

## 6　DISCUSSION AND CONCLUSIONS

This paper investigated the interaction within the social network of stakeholders in the requirements engineering phase of a public sector information system project. The project was analyzed with a view to determine the impact of social ties and knowledge sharing on the successful completion of the RE. The analysis showed that the social network and the knowledge drawn from there is crucial for the RE process. Particularly the importance of transactive memory was apparent in our data. Additionally, through our analysis we discovered an additional factor that had a great importance to a successful RE process, namely flexibility. It was evident from our data that flexibility in terms of making compromises to reach an acceptable solution for all parties was considered a very important factor affecting a successful outcome of the RE process. Interestingly, also the parties who had to be flexible in terms of their requirements indicated satisfaction with the overall outcome. It seemed to be more important for all parties to achieve a satisfactory outcome that will fulfill its purpose rather than insisting to have all their requirements fully included.

Our research results contribute to both research and practice. The effect of social ties and knowledge sharing in the RE process has clearly been an under researched area. Thus, our study makes a contribution to the RE literature and may open a new stream of research that future research can pursue. Our results also contribute to the general project management literature by offering insights applicable in the project initiation and requirements gathering phases. Our study also contributes to the team work literature. Our perspective emphasizing the social network was insightful since the requirements engineering team of the case project was not static, but dynamically changing over time. The social network studied had similar elements to virtual teams (e.g. Kotlarsky and Oshri 2005), which may explain the findings that the product success was more often mentioned than personal satisfaction. Because of the dynamic nature of the team there were less face to face meetings with all the participants. Moreover, practitioners, particularly project managers and teams, should find our results interesting. Recognizing that RE is a learning process and requires knowledge from various stakeholders would help project professionals reflect on their own practices. Our study draws the attention to the wider social network where knowledge necessary in the RE process can be acquired from.

The study has some limitations. It is based on a single case study in the public sector, which may be considered a rather specific context. At the same time, being able to provide insights on a rather unique case can also be considered a strength for the study. Arguably, our results on the importance of the social networks in the RE process may apply to other contexts as well, although one might expect there to be differences in the relative importance of the different factors, such as social ties, knowledge sharing and flexibility, depending on the specific context. We believe that our results can be generalized to similar contexts as in this study but further research is needed in different organizational contexts.

A very interesting avenue for future research would thus be repeating the research in other organizations and in different industry contexts. Another interesting area for future research would be extending the scope of research to cover the entire project life-cycle to investigate the effects of the social networks in the project execution phase and even in the post-adoption phase of information systems.

# Copyright